\documentclass[aps, prd, superscriptaddress, nofootinbib, showpacs, preprintnumbers, twocolumn]{revtex4}

\usepackage{graphicx}
\usepackage{amsmath}
\usepackage{amssymb}

\def\fsl#1{\setbox0=\hbox{$#1$}                 % set a box for #1
   \dimen0=\wd0                                 % and get its size
   \setbox1=\hbox{/} \dimen1=\wd1               % get size of /
   \ifdim\dimen0>\dimen1                        % #1 is bigger
      \rlap{\hbox to \dimen0{\hfil/\hfil}}      % so center / in box
      #1                                        % and print #1
   \else                                        % / is bigger
      \rlap{\hbox to \dimen1{\hfil$#1$\hfil}}   % so center #1
      /                                         % and print /
   \fi}                                         %

\newcommand{\NDA}{\Omega_{\rm NDA}}

\newcommand{\Ly}{\Lambda_{LY}}

\newcommand{\DLR}{\stackrel{\leftrightarrow}\partial}

\makeatletter
\@addtoreset{equation}{section}
\makeatother

\begin{document}
\title{Topcolor model in extra dimensions and nontrivial boundary 
conditions\footnote{Based on talk given at 
{\it The International Workshop on Dynamical Symmetry Breaking}, 
Nagoya University, Nagoya, Japan, 21-22 December 2004.}}
\date{\today}

\author{Michio Hashimoto}
\email[E-mail: ]{mhashimo@uwo.ca}
\affiliation{Department of Applied Mathematics,
Middlesex College, The University of Western Ontario,
London, Ontario N6A 5B7, Canada}

\begin{abstract}
The nontrivial boundary conditions for the Topcolor breaking 
are investigated in the context of the TeV-scale extra dimension scenario. 
We present a six dimensional model where the top and bottom quarks in 
the bulk have the Topcolor charge while the other quarks in the bulk do not. 
We also put the electroweak gauge interaction in the six dimensional bulk. 
Then the bottom quark condensation is naturally suppressed owing to 
the power-like running of the bulk $U(1)_Y$ interaction, so that 
only the top condensation is expected to take place. 
We explore such a possibility based on the ladder Schwinger-Dyson equation 
and show the cutoff to make the model viable.
\end{abstract}

\maketitle

\section{Introduction}

In the context of the TeV-scale extra dimension 
scenario~\cite{Antoniadis:1990ew,Dienes:1998vh}, 
the top quark condensate~\cite{MTY89,Nambu89,Marciano89,Bardeen:1989ds}
has been reconsidered by several authors~\cite{Dobrescu:1998dg,Cheng:1999bg,Kobakhidze:1999ce,Arkani-Hamed:2000hv,Hashimoto:2000uk,Gusynin:2002cu,Hashimoto:2003ve,Gusynin:2004jp}.
In particular, Arkani-Hamed, Cheng, Dobrescu and
Hall (ACDH)~\cite{Arkani-Hamed:2000hv} proposed a version of 
the top condensate model where the third generation quarks and leptons 
as well as the the Standard Model (SM) gauge bosons are put in the bulk,
while any four-fermion interactions are not introduced in the bulk.
In Refs.~\cite{Hashimoto:2000uk,Gusynin:2002cu},
the full bulk gauge dynamics was investigated,
based on the ladder Schwinger-Dyson (SD) equation.
The phenomenological implications
were studied in Ref.~\cite{Hashimoto:2003ve}.
It is found that the model with $D=8$ can be viable and
both masses of the top quark and Higgs boson are
predicted as $m_t=172$--$175$ GeV and $m_H=176$--$188$ GeV, respectively.
However it turns out that
{\it the simplest scenario with $D=6$ does not work: 
The tau condensation is favoured instead of the top.}

On the other hand, it is known that 
field theories in six dimensions have several interesting features
relating to proton stability~\cite{Appelquist:2001mj},
explanation of the number of the generations of 
fermions~\cite{Dobrescu:2001ae}, etc..
In order to construct a viable top condensate model in six dimensions,
a strong interaction other than the six dimensional QCD should be required,
because the attractive force provided by the bulk QCD is not sufficient 
to generate the top condensate.
A candidate for such a strong interaction is 
Topcolor.~\cite{Hill:1991at,TC2,Dobrescu:1997nm}
(See for reviews Refs.~\cite{Cvetic:1997eb,Hill:2002ap}. )

We may introduce the Topcolor interaction in the bulk.
Topcolor should be broken down in low energy.
In four dimensions, some involved dynamical mechanism is needed
in order to break Topcolor,
unless a (composite) scalar field is introduced for simplicity.
As for the gauge symmetry breaking,
the extra dimension scenario has an advantage:
The gauge symmetry breaking can be easily achieved
by imposing appropriate boundary conditions 
(BC's).~\cite{Kawamura:1999nj}
On the basis of more general BC's,
the Higgsless theory was proposed~\cite{Csaki:2003dt}.
The gauge symmetry breaking mechanism via 
nontrivial BC's can be also applied to other models for the dynamical 
electroweak symmetry breaking. 
We here study nontrivial BC's for the Topcolor 
breaking along with Ref.~\cite{Hashimoto:2004xz}.

Benefits of the Topcolor breaking through nontrivial BC's are
as follows:
The mechanism is very simple and hence it is easy to extend to
various models.
We can break spontaneously the Topcolor gauge symmetry without 
introducing explicitly a (composite) scalar field and thereby
we can carry out the model building incorporating only fermions and 
gauge bosons (in the bulk).
In passing, the Topcolor gauge bosons do not have mass terms in the bulk
in the gauge breaking mechanism via the BC's.
Therefore the theory does not provide four-fermion (NJL-type) interactions 
in the bulk, unlike four dimensional Topcolor models.

Let us investigate a six dimensional Topcolor model with 
the $SU(3)_1 \times SU(3)_2$ gauge symmetry.
We assign the Topcolor charge, $SU(3)_1$, to the top and bottom quarks 
in the bulk, while we do the $SU(3)_2$ charge
to the first and second generation quarks in the bulk.
We then impose the nontrivial BC's so that
$SU(3)_1 \times SU(3)_2$ breaks down to the diagonal subgroup,
which is identified to QCD.
We also put the electroweak gauge interaction in the bulk
and hence the electroweak gauge sector is the same as the universal 
extra dimension model~\cite{Appelquist:2000nn}.
In order to obtain the chiral fermion in four dimensions,
we apply the compactification on a square proposed by Dobrescu and
Pont\'{o}n~\cite{Dobrescu:2004zi}, which is closely related to
the compactification on the orbifold $T^2/Z_4$.

For a viable model it is required that only the top condensation occurs
while other condensations such as bottom and leptons do not.
The up and charm condensations should be suppressed as well.
We call the requirement ``tMAC condition'' and the energy scale
``tMAC scale'' as in Ref.~\cite{Hashimoto:2003ve}.
Once we specify the model, the renormalization group (RG) flows of
the gauge couplings can be determined through the truncated
Kaluza-Klein (KK) effective theory~\cite{Dienes:1998vh}. 
The running effects are very important to study the tMAC scale.
Suppression of the up and charm condensations can be realized by
the difference of the gauge coupling strengths between $SU(3)_1$ and 
$SU(3)_2$. Since the non-Abelian gauge theory in the bulk has 
the ultraviolet fixed point (UVFP) within the truncated KK effective 
theory~\cite{Hashimoto:2000uk}, the difference is essentially determined 
by the values of the UVFP for $SU(3)_1$ and $SU(3)_2$.
Note that the value of the UVFP is controlled by the number $N_f$ of 
fermions, i.e., the model parameter, and that the value increases 
as the number $N_f$ is larger.\footnote{The UVFP disappears above 
a certain number of fermions.}
In our model, we assign the $SU(3)_2$ charge to the first and
second generation quarks, i.e., $N_f=4$.
Therefore we need to introduce more than three (vector-like heavy) 
fermions with the $SU(3)_1$ charge other than the top and bottom quarks.
We then find that the top condensation takes place while the up and 
charm condensations are actually suppressed.
Can we realize the mass hierarchy between the top and bottom?
We here note that the bulk hypercharge interaction $U(1)_Y$
rapidly becomes strong owing to the power-like running.
Thus the $U(1)$ tilting mechanism to suppress the bottom quark condensation
is automatically incorporated in our model.
However the lepton condensation is apparently favoured extremely near 
the Landau pole of $U(1)_Y$.
It means that the tMAC scale will be found in the region that 
it is large enough to suppress the bottom condensation while 
it is smaller than the scale for the lepton condensation.
We concretely analyze the tMAC scale by using the ladder SD equation
and depict the results in two dimensional plane of the cutoff $\Lambda$
and the ratio of the Topcolor and QCD couplings $g^2(R^{-1})/g_3^2(R^{-1})$
at the compactification scale $R^{-1}$.
For a slice $g^2(R^{-1})/g_3^2(R^{-1})=4.6$, for example,
we find that the tMAC scale is $\Lambda R \sim 10$--$10.5$.
We also show that the model is not excluded by constraints of 
S, T-parameters.

The paper is organized as follows:
In Sec.~\ref{sec2} we study the chiral compactification on a square.
In Sec.~\ref{sec3} we investigate appropriate BC's for the Topcolor breaking.
In Sec.~\ref{sec4} we present the model and study the running
effects of the gauge couplings.
In Sec.~\ref{sec5} we determine the tMAC scale by solving
the ladder SD equation.
Sec.~\ref{summary} is devoted to summary and discussions.

\section{Chiral compactification on a square}
\label{sec2}

Let us study six dimensional gauge theories with chiral fermions.
We compactify extra two spatial dimensions $(y^5,y^6)$
on a square with $0 \leq y^5, y^6 \leq L$.
Since chiral fermions in six dimensions contain both of
right and left handed components as the four dimensional chirality,
we must carry out a chiral compactification in order to keep 
only the zero modes identified to the conventional SM particles.

For a while, we consider a gauge theory with a chiral fermion $\psi_+^{}$ 
in the bulk,
\begin{equation}
  {\cal L} = {\cal L}_{\psi_+^{}} + {\cal L}_{\rm gauge} ,
\end{equation}
with
\begin{equation}
  {\cal L}_{\psi_+^{}} = \bar{\psi}_+ iD_M \Gamma^M \psi_+ ,
\end{equation}
and
\begin{equation}
  {\cal L}_{\rm gauge} =  - \frac{1}{4}F_{MN}^a F^{a\,MN} , \label{L_g}
\end{equation}
where $M,N=0,1,2,3,5,6$, and $\Gamma^M$ denotes the gamma matrices
in six dimensions.
We here defined
\begin{equation}
  D_M \equiv \frac{1}{2}\DLR_M - i g_{6D}^{} A_M,
\end{equation}
\begin{equation}
  \bar{\psi}\!\DLR_M \!\!\Gamma^M \psi \equiv
  \bar{\psi} \Gamma^M (\partial_M \psi)
 -(\partial_M \bar{\psi}) \Gamma^M \psi ,
\end{equation}
and
\begin{equation}
  F_{MN}^a \equiv \partial_M A_N^a - \partial_N A_M^a
  + g_{6D}^{} f^{abc} A_M^b A_N^c ,
\end{equation}
where $f^{abc}$ is the structure constant of the gauge group,
$g_{6D}^{}$ the {\it dimensionful} bulk gauge coupling constant.
The chiral fermions $\psi_\pm$ in the bulk are defined as
\begin{equation}
  \psi_\pm \equiv P_\pm \psi,
\end{equation}
with the chiral projection operators $P_\pm$,
\begin{equation}
  P_\pm \equiv \frac{1}{2}\left(1 \pm \Gamma_{\chi,7} \right),
\end{equation}
where the chirality matrix $\Gamma_{\chi,7}$ in six dimensions is
\begin{equation}
  \Gamma_{\chi,7}\equiv
  \Gamma^0 \Gamma^1 \Gamma^2 \Gamma^3 \Gamma^5 \Gamma^6, \quad
  \Gamma_{\chi,7} \Gamma_{\chi,7} = 1.
\end{equation}
It is straightforward to incorporate $\psi_-^{}$ and 
other gauge bosons, $A'_M,\cdots$.

For our purpose, it is convenient to use four dimensional
right/left-handed notations.
The four dimensional chirality matrix $\Gamma_{\chi,5}$ is defined by
\begin{equation}
 \Gamma_{\chi,5}\equiv i \Gamma^0 \Gamma^1 \Gamma^2 \Gamma^3 ,\quad
 \Gamma_{\chi,5} \Gamma_{\chi,5}=1.
\end{equation}
The matrices $\Gamma_{\chi,5}$ and $\Gamma_{\chi,7}$ satisfy
\begin{equation}
 [\Gamma_{\chi,5},\Gamma_{\chi,7}] = 0,
\end{equation}
so that $\Gamma_{\chi,5}$ and $\Gamma_{\chi,7}$ are simultaneously
diagonalizable.
Thus we further decompose $\psi_\pm$ into the four dimensional
right/left-handed fermions:
\begin{equation}
  \psi_\pm = \psi_{\pm R} + \psi_{\pm L},
\end{equation}
where
\begin{equation}
  \psi_{\pm R} \equiv P_R \psi_{\pm}, \quad
  \psi_{\pm L} \equiv P_L \psi_{\pm},
\end{equation}
with the four dimensional chiral projection operators $P_{R,L}$,
\begin{equation}
  P_{R,L} \equiv \frac{1}{2}\left(1 \pm \Gamma_{\chi,5} \right).
\end{equation}
Noting
\begin{equation}
  \{\Gamma^\mu,\Gamma_{\chi,5}\}=0, \quad \mbox{for} \quad
  \mu = 0,1,2,3
\end{equation}
and
\begin{equation}
  [\Gamma^m,\Gamma_{\chi,5}]=0, \quad \mbox{for} \quad
  m=5,6,
\end{equation}
the Lagrangian ${\cal L}_{\psi_+^{}}$ is rewritten in terms of
$\psi_{+R}$ and $\psi_{+L}$ as follows:
\begin{equation}
  {\cal L}_{\psi_+^{}} = {\cal L}_{RR+LL}^{} + {\cal L}_{RL+LR}^{},
\end{equation}
with
\begin{equation}
{\cal L}_{RR+LL}^{}  \equiv
  \bar{\psi}_{+R} iD_\mu \Gamma^\mu \psi_{+R}
+ \bar{\psi}_{+L} iD_\mu \Gamma^\mu \psi_{+L},
  \label{Lag-psi-RR}
\end{equation}
and
\begin{equation}
{\cal L}_{RL+LR}^{} \equiv
  \bar{\psi}_{+R} iD_m \Gamma^m \psi_{+L}
+ \bar{\psi}_{+L} iD_m \Gamma^m \psi_{+R}.
  \label{Lag-psi-RL}
\end{equation}

Following Dobrescu and Pont\'{o}n~\cite{Dobrescu:2004zi},
we identify two adjacent sides as follows:
\begin{equation}
  (y,0) \equiv (0,y), \quad (y,L) \equiv (L,y), \quad
  {}^\forall \!y \in [0,L] , \label{z4}
\end{equation}
which is closely related to the orbifold compactification on $T^2/Z_4$.
Under the identification (\ref{z4}), the Lagrangian should be the same:
\begin{equation}
  {\cal L}|^{(y,0)} = {\cal L}|^{(0,y)}, \quad
  {\cal L}|^{(y,L)} = {\cal L}|^{(L,y)}. \label{Lag_z4}
\end{equation}
We then impose the BC's on fermions as
\begin{subequations}
\label{f-RL-z4}
\begin{align}
  \psi_{+R}(y,0) &= e^{\frac{i\pi}{2}n}\psi_{+R}(0,y),
  \label{f-R-z4}
  \\
  \psi_{+L}(y,0) &= ie^{\frac{i\pi}{2}n}\psi_{+L}(0,y),
  \label{f-L-z4}
\end{align}
\end{subequations}
and
\begin{subequations}
\label{f-RL2-z4}
\begin{align}
  \psi_{+R}(y,L) &= (-1)^\ell e^{\frac{i\pi}{2}n}\psi_{+R}(L,y),
  \label{f-R2-z4}
  \\
  \psi_{+L}(y,L) &= i\,(-1)^\ell e^{\frac{i\pi}{2}n}\psi_{+L}(L,y),
  \label{f-L2-z4}
\end{align}
\end{subequations}
where the integers $n$ and $\ell$ can take the values of
$n=0,1,2,3$ and $\ell =0,1$, respectively.
Differentiating the BC's (\ref{f-RL-z4})--(\ref{f-RL2-z4})
with respect to $y$, we find
\begin{subequations}
\label{df-RL-z4}
\begin{align}
  \partial_5 \psi_{+R}(y,0) &=
  e^{\frac{i\pi}{2}n} \partial_6 \psi_{+R}(0,y),
  \label{df-R-z4}
  \\
  \partial_5 \psi_{+L}(y,0) &=
  ie^{\frac{i\pi}{2}n} \partial_6 \psi_{+L}(0,y),
  \label{df-L-z4}
\end{align}
\end{subequations}
and
\begin{subequations}
\label{df-RL2-z4}
\begin{align}
  \partial_5 \psi_{+R}(y,L) &=
  (-1)^\ell e^{\frac{i\pi}{2}n} \partial_6 \psi_{+R}(L,y),
  \label{df-R2-z4}
  \\
  \partial_5 \psi_{+L}(y,L) &=
  i\,(-1)^\ell e^{\frac{i\pi}{2}n} \partial_6 \psi_{+L}(L,y).
  \label{df-L2-z4}
\end{align}
\end{subequations}
We further impose the BC's on the derivative terms as
\begin{subequations}
\label{df2-RL-z4}
\begin{align}
  \partial_6 \psi_{+R}(y,0) &=
  -e^{\frac{i\pi}{2}n} \partial_5 \psi_{+R}(0,y),
  \label{df2-R-z4}
  \\
  \partial_6 \psi_{+L}(y,0) &=
  -ie^{\frac{i\pi}{2}n} \partial_5 \psi_{+L}(0,y),
  \label{df2-L-z4}
\end{align}
\end{subequations}
and
\begin{subequations}
\label{df2-RL2-z4}
\begin{align}
  \partial_6 \psi_{+R}(y,L) &=
  (-1)^{\ell+1} e^{\frac{i\pi}{2}n} \partial_5 \psi_{+R}(L,y),
  \label{df2-R2-z4}
  \\
  \partial_6 \psi_{+L}(y,L) &=
  i\,(-1)^{\ell+1} e^{\frac{i\pi}{2}n} \partial_5 \psi_{+L}(L,y).
  \label{df2-L2-z4}
\end{align}
\end{subequations}
The BC's of the derivative terms imply the identification of
gauge bosons as
\begin{subequations}
\label{gauge-z4}
\begin{align}
  A_\mu (y,0) &= A_\mu (0,y), & A_\mu (y,L) &= A_\mu (L,y),
  \label{g-z4}
  \\
  A_{5} (y,0) &= A_{6}(0,y),  & A_{5} (y,L) &= A_{6}(L,y),
  \label{g5-z4}
  \\
  A_{6} (y,0) &= -A_{5}(0,y), & A_{6} (y,L) &= -A_{5}(L,y).
  \label{g6-z4}
\end{align}
\end{subequations}
We differentiate Eq.~(\ref{gauge-z4}) with respect to $y$ and find
\begin{subequations}
\label{dgauge-z4}
\begin{eqnarray}
  \partial_5 A_\mu |^{(y,0),\,(y,L)} &=& \partial_6 A_\mu |^{(0,y),\,(L,y)},
  \label{dg-z4}
  \\[2mm]
  \partial_5 A_{6}|^{(y,0),\,(y,L)} &=& - \partial_6 A_{5}|^{(0,y),\,(L,y)}.
  \label{dg6-z4}
\end{eqnarray}
\end{subequations}
The identification (\ref{Lag_z4}) for the gauge sector ${\cal L}_{\rm gauge}$
requires the BC's
\begin{subequations}
\label{dgauge2-z4}
\begin{eqnarray}
  \partial_6 A_\mu |^{(y,0),\,(y,L)} &=& -\partial_5 A_\mu |^{(0,y),\,(L,y)},
  \label{dg2-z4}
  \\[2mm]
  \partial_6 A_{5}|^{(y,0),\,(y,L)} &=& - \partial_5 A_{6}|^{(0,y),\,(L,y)}.
  \label{dg5-z4}
\end{eqnarray}
\end{subequations}

Now it is easy to check that the identification (\ref{Lag_z4}) is
satisfied. From the BC's (\ref{f-RL-z4})--(\ref{f-RL2-z4}),
${\cal L}_{RR+LL}^{}$ defined by Eq.~(\ref{Lag-psi-RR})
is obviously identical to the reflection under Eq.~(\ref{z4}).
To see the identity for ${\cal L}_{RL+LR}^{}$,
we apply the relations
\begin{equation}
  \Gamma^5 P_R P_\pm = \pm i\Gamma^6 P_R P_\pm, \quad
  \Gamma^5 P_L P_\pm = \mp i\Gamma^6 P_L P_\pm,
  \label{gamma5-6}
\end{equation}
and thereby rewrite ${\cal L}_{RL+LR}^{}$ in Eq.~(\ref{Lag-psi-RL}) as
\begin{eqnarray}
{\cal L}_{RL+LR}^{} &=& \phantom{+}
  \bar{\psi}_{+R} (D_5\Gamma^6 - D_6 \Gamma^5) \psi_{+L}
  \nonumber \\ && +
  \bar{\psi}_{+L} (-D_5\Gamma^6 + D_6 \Gamma^5) \psi_{+R}.
  \label{Lag-psi-RL2}
\end{eqnarray}
By using the BC's of Eqs.~(\ref{f-RL-z4})--(\ref{df2-RL2-z4}) and 
the representation of ${\cal L}_{RL+LR}^{}$ in Eq.~(\ref{Lag-psi-RL2}),
we can also confirm the identity
${\cal L}_{RL+LR}^{}\,|^{(y,0),\,(y,L)} =
 {\cal L}_{RL+LR}^{}\,|^{(0,y),\,(L,y)}$.
How about the identity for the gauge sector?
The derivative of Eq.~(\ref{gauge-z4}) with respect to $x^\mu$ and
Eqs.~(\ref{dgauge-z4})--(\ref{dgauge2-z4}) yield
\begin{subequations}
\label{F-z4}
\begin{eqnarray}
 F_{\mu\nu}^a\,|^{(y,0),\,(y,L)} &=& \phantom{-}
 F_{\mu\nu}^a\,|^{(0,y),\,(L,y)},\\
 F_{\mu 5}^a\,|^{(y,0),\,(y,L)} &=& \phantom{-}
 F_{\mu 6}^a\,|^{(0,y),\,(L,y)},\\
 F_{\mu 6}^a\,|^{(y,0),\,(y,L)} &=& 
 -F_{\mu 5}^a\,|^{(0,y),\,(L,y)}, \label{F_mu6} \\
 F_{56}^a\,|^{(y,0),\,(y,L)} &=& \phantom{-}
 F_{56}^a\,|^{(0,y),\,(L,y)},
\end{eqnarray}
\end{subequations}
so that the identity ${\cal L}_{\rm gauge}\,|^{(y,0),\,(y,L)} =
{\cal L}_{\rm gauge}\,|^{(0,y),\,(L,y)}$ is clearly satisfied.

Can be the one of the zero modes really moved away?
To explain this concretely, we take $n=0$, $\ell=0$ in 
Eqs.~(\ref{f-RL-z4})--(\ref{f-RL2-z4}).
Since the zero mode does not depend on the extra spacial coordinates 
$y^5$ and $y^6$ by definition, the left handed part $\psi_{+L}$ cannot 
include any zero modes. 
On the other hand, the right handed part $\psi_{+R}$ does include
a zero mode consistently with the BC's (\ref{f-RL-z4})--(\ref{df2-RL2-z4}).
When we take $n=3,\ell=0$, $\psi_{+L}$ has a zero mode 
while $\psi_{+R}$ does not.
In this way, we can achieve the chiral compactification through 
the identification (\ref{z4}).

\section{Nontrivial boundary conditions for the Topcolor Breaking}
\label{sec3}

We investigate appropriate BC's for the Topcolor breaking.

First, we start from the Lagrangian (\ref{L_g}) with a single gauge symmetry
and discuss nontrivial BC's with or without the gauge symmetry breaking.
Next, we extend the results to the Topcolor model with the
$SU(3)_1 \times SU(3)_2$ gauge symmetry.
Finally, we include the top quark in the bulk.

\subsection{Nontrivial boundary conditions}

After integration by parts the variation of the action with respect to
the gauge field yields the equation of motion (EOM) and 
the condition~\cite{Hashimoto:2004xz},
\begin{equation}
 F_{5\mu}^a\,\delta\! A^{a\,\mu}\bigg|_{(0,y)}^{(L,y)}
+F_{6\mu}^a\, \delta\! A^{a\,\mu}\bigg|_{(y,0)}^{(y,L)} =0, \label{dS}
\end{equation}
where
\begin{equation}
  X \bigg|_{(0,y)}^{(L,y)} \equiv X(x^\mu,L,y)-X(x^\mu,0,y) , \label{X}
\end{equation}
and similar is the definition of $X |_{(y,0)}^{(y,L)}$.
Note that Eq.~(\ref{dS}) is always satisfied under the chiral
compactification owing to Eqs.~(\ref{g-z4}) and (\ref{F_mu6}).

We may impose  (i) the Neumann-Neumann (NN) BC's on the gauge vector 
field $A_\mu$,
\begin{equation}
 \partial_5 A_\mu^a |^{(0,y),(L,y)} = 0, \qquad
 \partial_6 A_\mu^a |^{(y,0),(y,L)} = 0,
 \label{bc-NN}
\end{equation}
or (ii) the Dirichlet-Dirichlet (DD) BC's,
\begin{equation}
 A_\mu^a (0,y) = A_\mu^a (L,y) = A_\mu^a (y,0) = A_\mu^a (y,L) = 0 .
 \label{bc-DD}
\end{equation}
Some of BC's for $A_\mu$ such as the Neumann-Dirichlet (ND), 
$\partial_5 A_\mu^a |^{(0,y),(L,y)}=0, A_\mu^a|^{(y,0),(y,L)}=0$, 
disagree with the requirements of the chiral compactification,
Eqs.~(\ref{gauge-z4})--(\ref{dgauge2-z4}).
We henceforth consider only the cases of (i) and (ii).
It is obvious that the NN BC's (i) respect the four dimensional gauge 
symmetry, while the DD BC's (ii) break the gauge symmetry.

What kind of BC's is appropriate for the gauge scalars $A_{5,6}$?
For example, BC's such as $ A_{5,6}^a |^{(0,y),(L,y),(y,0),(y,L)} = 0$ is
inconsistent with the gauge symmetry.
We find the following nontrivial BC's 
consistent with the gauge symmetry and the EOM:
\begin{displaymath}
 \mbox{(i) NN for $A_\mu$}  \quad \to \quad 
 \mbox{DN for $A_5$ and ND for $A_6$}
\end{displaymath}
\begin{eqnarray}
 \partial_5 A_\mu^a |^{(0,y),(L,y)} = 0, &&
 \partial_6 A_\mu^a |^{(y,0),(y,L)} = 0, \\
 A_5^a |^{(0,y),(L,y)} = 0,  && \partial_6 A_5^a |^{(y,0),(y,L)} = 0,\\
 \partial_5 A_6^a |^{(0,y),(L,y)} = 0, &&
 \phantom{\partial_6} A_6^a |^{(y,0),(y,L)} = 0 ,
\end{eqnarray}
and
\begin{displaymath}
 \mbox{(ii) DD for $A_\mu$}  \quad \to \quad 
 \mbox{ND for $A_5$ and DN for $A_6$}
\end{displaymath}
\begin{eqnarray}
  A_\mu^a |^{(0,y),(L,y)} = 0, &&
 \phantom{\partial_6}A_\mu^a |^{(y,0),(y,L)} = 0, \\
 \partial_5 A_5^a |^{(0,y),(L,y)} = 0, &&
 \phantom{\partial_6}A_5^a |^{(y,0),(y,L)} = 0, \\ 
 A_6^a |^{(0,y),(L,y)} = 0, && \partial_6 A_6^a |^{(y,0),(y,L)} = 0 .
\end{eqnarray}
These BC's are also consistent with the chiral compactification.
Under the BC's (i), the system has the five dimensional gauge symmetry
on each side.
A remarkable point is that gauge scalars do not have zero modes
in both cases (i) and (ii).
We here note that one of the gauge scalars can be identically zero 
by taking the unitary gauge.

\subsection{Topcolor breaking}

Let us analyze the $SU(3)_1 \times SU(3)_2$ gauge theory in the bulk.
The Lagrangian is given by
\begin{equation}
  {\cal L}_g =
  - \frac{1}{4}F_{MN}^a F^{a\,MN} - \frac{1}{4}F_{MN}^{'a} F^{'a\,MN},
\end{equation}
with
\begin{equation}
  F_{MN}^{'a} \equiv \partial_M A_N^{'a} - \partial_N A_M^{'a}
  + g'_{6D} f^{abc} A_M^{'b} A_N^{'c}. \label{F}
\end{equation}
The gauge fields $A_M^a$ and $A_M^{'a}$ are associated with the gauge groups
$SU(3)_1$ and $SU(3)_2$, respectively.
We assign the Topcolor to the $SU(3)_1$ gauge interaction.

We break the gauge symmetry $SU(3)_1 \times SU(3)_2$ to the diagonal subgroup
by assigning appropriate BC's to the gauge fields.
The unbroken subgroup is identified to the conventional QCD.
For such a purpose, we may choose the BC's (i) for the 
``gluon'' fields and the BC's (ii) for the ``coloron'' fields.

We now define the ``gluon'' field $G_M$ and
the ``coloron'' field $G'_M$ as
\begin{equation}
 \left\{
  \begin{array}{lcl}
  G_M(x^\mu,y^5,y^6) & = & \phantom{+}
  A'_M \cos\theta + A_M \sin\theta , \\[2mm]
  G'_M(x^\mu,y^5,y^6) & = &
  - A'_M \sin\theta + A_M \cos\theta ,
  \end{array}
 \right. \label{G-A}
\end{equation}
where $\theta$ denotes a ``mixing angle'' and we used the notation
\begin{equation}
  A_M \equiv A_M^a T^a,
\end{equation}
with $T^a$ being the generator of the SU(3) Lie algebra.
In order to realize the Topcolor breaking, 
we assign the following BC's to $G_M$ and $G'_M$:
\begin{eqnarray}
 \partial_5 G_\mu |^{(0,y),(L,y)} = 0, &&
 \partial_6 G_\mu |^{(y,0),(y,L)} = 0, \\
 G_5 |^{(0,y),(L,y)} = 0,  && \partial_6 G_5 |^{(y,0),(y,L)} = 0,\\
 \partial_5 G_6 |^{(0,y),(L,y)} = 0, &&
 \phantom{\partial_6} G_6 |^{(y,0),(y,L)} = 0 ,
\end{eqnarray}
and
\begin{eqnarray}
  G'_\mu |^{(0,y),(L,y)} = 0, &&
 \phantom{\partial_6}G'_\mu |^{(y,0),(y,L)} = 0, \\
 \partial_5 G'_5 |^{(0,y),(L,y)} = 0, &&
 \phantom{\partial_6}G'_5 |^{(y,0),(y,L)} = 0, \\ 
 G'_6 |^{(0,y),(L,y)} = 0, && \partial_6 G'_6 |^{(y,0),(y,L)} = 0 .
\end{eqnarray}
The corresponding BC's for $A_M$ and $A'_M$ read from Eq.~(\ref{G-A}),
i.e.,
\begin{equation}
 A_\mu |^{(0,y),(L,y),(y,0),(y,L)} =
 \tan\theta \,A'_\mu|^{(0,y),(L,y),(y,0),(y,L)},
 \label{bc-A}
\end{equation}
etc..

\subsection{Topcolor model on a square}

Let us take into account the top quark $T$ in the bulk,
which has the $SU(3)_1$ charge,
\begin{equation}
  {\cal L}_t = \bar{T}_+\,i D_M \Gamma^M T_+
 + \bar{T}_- \,i D_M \Gamma^M T_- .
\end{equation}
We take a notation that $T_{+R,-L}$ include
the SM-like top quarks $t_{R,L}$ as the zero modes.
It corresponds to the convention of $n=0$ and $\ell=0$
in Eqs.~(\ref{f-RL-z4})--(\ref{df2-RL2-z4}).
The desirable BC's for $T_{+R}$ and $T_{-L}$ are given by
\begin{equation}
 \partial_5 T_{+R,-L}|^{(0,y),(L,y)} = 0 , \quad
 \partial_6 T_{+R,-L}|^{(y,0),(y,L)} = 0 .
 \label{bc-dT}
\end{equation}
The heavy components $T_{+L}$ and $T_{-R}$ are determined by 
consistency with the EOM. For details, see 
Refs.~\cite{Dobrescu:2004zi,Hashimoto:2004xz}.

The KK decompositions of $T_{+R,-L}$, $G_\mu$ and $G'_\mu$ are obtained as
\begin{equation}
  T_{+R,-L}(x^\mu,y^5,y^6) =
  \frac{1}{L} \sum_{j \geq k \geq 0} T_{+R,-L}^{[j, k]}(x^\mu)
  f_{cc}^{[j,k]}(y^5,y^6) ,
\label{tR}
\end{equation}
\begin{equation}
  G_\mu(x^\mu,y^5,y^6) =
  \frac{1}{L} \sum_{j \geq k \geq 0} G_{\mu}^{[j, k]}(x^\mu)
  f_{cc}^{[j,k]}(y^5,y^6), \label{Gkk}
\end{equation}
\begin{equation}
  G'_\mu(x^\mu,y^5,y^6) =
  \frac{1}{L} \sum_{j > k > 0} G_{\mu}^{'\,[j, k]}(x^\mu)
  f_{ss}^{[j,k]}(y^5,y^6), \label{Gpkk}
\end{equation}
with
\begin{eqnarray}
  f_{cc}^{[j,k]} &\equiv& {\cal N}_{cc}
  \left[\,
  \cos\left(\dfrac{\pi}{L} j y^5\right)
  \cos\left(\dfrac{\pi}{L} k y^6\right)
  \right.  \nonumber \\ && \qquad \left.
 +\cos\left(\dfrac{\pi}{L} k y^5\right)
  \cos\left(\dfrac{\pi}{L} j y^6\right) \,\right],
\end{eqnarray}
\begin{eqnarray}
  f_{ss}^{[j,k]} &\equiv& {\cal N}_{ss}
  \left[\,
  \sin\left(\dfrac{\pi}{L} j y^5\right)
  \sin\left(\dfrac{\pi}{L} k y^6\right)
  \right.  \nonumber \\ && \qquad \left.
 -\sin\left(\dfrac{\pi}{L} k y^5\right)
  \sin\left(\dfrac{\pi}{L} j y^6\right) \,\right] ,
\end{eqnarray}
where ${\cal N}_{cc}$ and ${\cal N}_{ss}$ are certain normalization factors.
(For details, see Ref.~\cite{Hashimoto:2004xz}.)
In particular, the function $f_{cc}^{[0,0]}$ for the zero mode is given by
\begin{equation}
  f_{cc}^{[0,0]} = 1.
\end{equation}

{}From the symmetry breaking pattern $SU(3)_1 \times SU(3)_2 \to SU(3)_c$,
the gauge couplings of $SU(3)_1$ and $SU(3)_2$ are related to 
the QCD coupling.
Integrating the six dimensional Lagrangian over $dy^5$ and $dy^6$,
we define the four dimensional theory,
\begin{equation}
  {\cal L}_{4D} \equiv \int_0^L dy^5 \int_0^L dy^6 {\cal L}_{6D} ,
  \label{4D-6D}
\end{equation}
with
\begin{equation}
  {\cal L}_{6D} = {\cal L}_t + {\cal L}_g.
\end{equation}
By using Eqs.~(\ref{tR})--(\ref{Gpkk}) and the definition (\ref{G-A}),
we find the interaction term between zero modes of the top and the gluon as
\begin{equation}
  {\cal L}_{\rm int} = \frac{g_{6D}\sin\theta}{L}
  \bar{T}_{+R,-L}^{[0,0]}\Gamma^\mu G_\mu^{[0,0]}
  T_{+R,-L}^{[0,0]} . \label{L-int}
\end{equation}
We here note that the definition (\ref{4D-6D}) implies the relations
between the six and four dimensional gauge couplings as
\begin{equation}
  g_{6D}^2 = L^2 g^2, \quad  g_{6D}^{'2} = L^2 g'{}^2, \label{gd}
\end{equation}
where $g$ and $g'$ denote the four dimensional gauge coupling constants
for $SU(3)_1$ and $SU(3)_2$, respectively.
Eq.~(\ref{L-int}) then yields the relation
\begin{equation}
  g_3 = g \sin\theta, \label{g-g3}
\end{equation}
where $g_3$ is the four dimensional QCD coupling.
Similarly, we will assign the $SU(3)_2$ charge to the up quark $U$ 
in the bulk, etc..
The interaction term between the zero modes and the gluon is given by
\begin{equation}
  {\cal L}_{\rm int} = \frac{g'_{6D}\cos\theta}{L}
  \bar{U}_{+R,-L}^{[0,0]}\Gamma^\mu G_\mu^{[0,0]}
  U_{+R,-L}^{[0,0]} . \label{L-int2}
\end{equation}
We thus obtain a similar relation between QCD and
$SU(3)_2$ couplings,
\begin{equation}
  g_3 = g' \cos\theta. \label{gp-g3}
\end{equation}
Eqs.~(\ref{g-g3}) and (\ref{gp-g3}) read
\begin{equation}
  \frac{1}{g_3^2} = \frac{1}{g^2}+\frac{1}{g^{'2}},
  \label{g3-g}
\end{equation}
and the ``mixing angle'' introduced in Eq.~(\ref{G-A}) 
must satisfies
\begin{equation}
  \tan \theta = \frac{g'}{g} = \frac{g'_{6D}}{g_{6D}}. 
\end{equation}

\section{The model}
\label{sec4}

\begin{table}[tb]
 \centering
  \begin{tabular}{|c|cccc|}\hline
   & $SU(3)_1$ & $SU(3)_2$ & $SU(2)_W$ & $U(1)_Y$ \\ \hline
   $(t,b)_-$ & \bf{3} & \bf{1} & \bf{2} & $1/6$ \\
   $t_+$ & \bf{3} & \bf{1} & \bf{1} & $2/3$ \\
   $b_+$ & \bf{3} & \bf{1} & \bf{1} & $-1/3$ \\
   $(\nu_\tau,\tau)_-$ & \bf{1} & \bf{1} & \bf{2} & $-1/2$ \\
   $\tau_+$ & \bf{1} & \bf{1} & \bf{1} & $-1$ \\ \hline \hline
   $(c,s)_-$ & \bf{1} & \bf{3} & \bf{2} & $1/6$ \\
   $c_+$ & \bf{1} & \bf{3} & \bf{1} & $2/3$ \\
   $s_+$ & \bf{1} & \bf{3} & \bf{1} & $-1/3$ \\
   $(\nu_\mu,\mu)_-$ & \bf{1} & \bf{1} & \bf{2} & $-1/2$ \\
   $\mu_+$ & \bf{1} & \bf{1} & \bf{1} & $-1$ \\ \hline \hline
   $(u,d)_-$ & \bf{1} & \bf{3} & \bf{2} & $1/6$ \\
   $u_+$ & \bf{1} & \bf{3} & \bf{1} & $2/3$ \\
   $d_+$ & \bf{1} & \bf{3} & \bf{1} & $-1/3$ \\
   $(\nu_e,e)_-$ & \bf{1} & \bf{1} & \bf{2} & $-1/2$ \\
   $e_+$ & \bf{1} & \bf{1} & \bf{1} & $-1$ \\ \hline \hline
   $\psi_X$ & \bf{3} & \bf{1} & \bf{1} & $0$ \\ \hline
  \end{tabular}
  \caption{The charge assignment of the model. \label{tab1}}
\end{table}

We now incorporate all quarks and leptons of the SM into the model.
We put all of gauge fields and SM fermions in the six dimensional bulk.
We may introduce right-handed neutrinos in the bulk, which are not
relevant in the following analysis.

Let us assign the $SU(3)_1$ charge to 
the bulk top and bottom quarks.
We put the $SU(3)_2$ charge on the first and second generation quarks 
in the bulk.
The electroweak gauge sector is taken to the same as the universal 
extra dimension model~\cite{Appelquist:2000nn}.
We perform the chiral compactification described in Sec.~\ref{sec2}.
The Topcolor interaction should be sufficiently strong to trigger
the top condensation, so that
we further introduce vector-like (heavy) fermions $\psi_X$
having the $SU(3)_1$ charge in order to adjust the RG flow of $SU(3)_1$.
We show the charge assignment in Table~\ref{tab1}.

While $SU(3)_1$ and $SU(3)_2$ are vector-like,
the $SU(2)_W$ and $U(1)_Y$ representations are chiral.
Although the six dimensional theory is anomalous under
the charge assignment in Table~\ref{tab1}, the anomalies can be cancelled
out by the Green-Schwarz mechanism~\cite{Green:sg}.
We assume that the Green-Schwarz counterterm does not change
the results in the following analysis.

Let us study running of gauge couplings in
the ``truncated KK'' effective theory~\cite{Dienes:1998vh}
based on the $\overline{\rm MS}$-scheme.
In this section, we use the unit of the extra momentum $R^{-1}$
instead of $L$,
\begin{equation}
  R^{-1} \equiv \frac{\pi}{L} .
\end{equation}
We expand bulk fields into KK modes and
construct a four dimensional effective theory.
Below $R^{-1}$
RGEs of four dimensional gauge couplings
$g_i (i=3,2,Y)$ are given by those of the SM,
\begin{equation}
  (4\pi)^2 \mu \frac{d g_i}{d \mu} = b_i\,g_i^3, \quad (\mu < R^{-1})
\end{equation}
with $b_3=-7, b_2=-\frac{19}{6}$ and $b_Y=\frac{41}{6}$.
Above $R^{-1}$
QCD should be replaced by
the $SU(3)_1 \times SU(3)_2$ gauge interaction.
We also need to take into account contributions
of KK modes in $\mu \geq R^{-1}$.
Since the KK modes heavier than the renormalization scale $\mu$
are decoupled in the $\overline{\rm MS}$-RGEs,
we only need summing up the loops of the KK modes lighter than $\mu$.
We estimate the total number of KK modes below $\mu$ by the volume
of the momentum space of extra dimensions dividing by the identification
factor $n$,
\begin{equation}
N_{\rm KK}(\mu) = \frac{\pi (\mu R)^2}{n} , \quad (\mu \gg R^{-1}) .
  \label{nkk_app}
\end{equation}
Note that we impose additional BC's such as Eq.~(\ref{bc-dT})
other than the BC's for the $T^2/Z_4$ compactification.
Therefore our model corresponds to the case of 
\begin{equation}
  n=8.
\end{equation}
The estimate (\ref{nkk_app}) works well for $\mu R \gg 1$.
(See, e.g. Ref.~\cite{Hashimoto:2003ve}.~)
Within the truncated KK effective theory, we obtain the RGE
\begin{equation}
  (4\pi)^2 \mu \frac{d g}{d \mu} =
  N_{\rm KK} (\mu) \,b_{\rm tc} \,g^3,
  \quad (\mu \geq R^{-1}) \label{rge_ED}
\end{equation}
with
\begin{equation}
  b_{\rm tc} = -\frac{22}{3} \, + \frac{4}{3} \cdot N_X, \quad
 \mbox{for} \quad SU(3)_1, \label{b3KK}
\end{equation}
where $N_X$ is the number of $\psi_X$ with the fundamental representation.
Other RGE coefficients are given by
\begin{align}
  b' &= -\frac{14}{3} , & {\rm for} & \quad SU(3)_2, \\
  b'_2 &= \frac{4}{3} + \frac{1}{6} \, n_h,  & {\rm for} & \quad SU(2)_W, \\
  b'_Y &= \frac{40}{3} + \frac{1}{6} \, n_h, & {\rm for} & \quad U(1)_Y.
\end{align}
In the following analysis, we assume that one composite Higgs doublet
appears in the low-energy spectrum, i.e., $n_h=1$.

We define the {\it dimensionless} bulk gauge coupling $\hat g$ as
$\hat g^2 \equiv g_{6D}^2 \mu^2$
and thereby obtain
\begin{equation}
  \hat g^2(\mu) = (\pi R \mu)^2 g^2 (\mu),
  \label{hat-g}
\end{equation}
where we used Eq.~(\ref{gd}).
Combining Eq.~(\ref{hat-g}) with the RGE~(\ref{rge_ED}),
we find RGEs for the dimensionless bulk Topcolor coupling $\hat g$,
\begin{equation}
 \mu \frac{d}{d \mu} \hat g = \hat g
 + \NDA\, b_{\rm tc}\, \hat g^3 , \label{rge_ED2}
\end{equation}
with $\NDA$ being the loop factor in $D$ dimensions,
\begin{equation}
  \NDA \equiv \frac{1}{(4\pi)^{D/2}\Gamma(D/2)} .
\end{equation}
The RGEs for $SU(3)_2$, $SU(2)_W$, and $U(1)_Y$ are the same as
Eq.~(\ref{rge_ED2}) with $b'$, $b'_2$, and $b'_Y$, respectively.

\begin{figure}[tbp]
  \begin{center}
    \resizebox{0.45\textwidth}{!}{
     \includegraphics{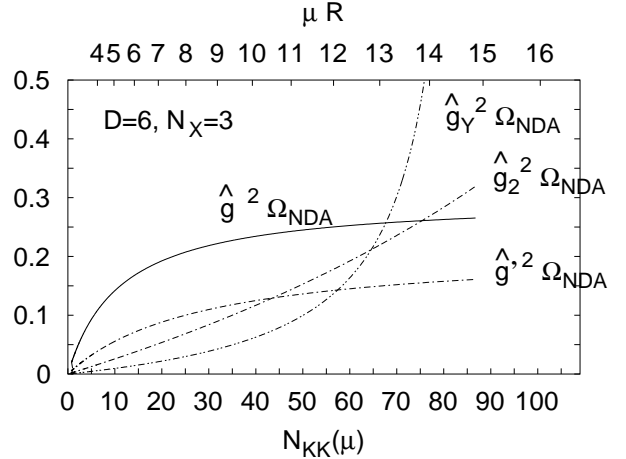}}
    \caption{Typical RG flows for the model with $N_X=3$.
             The ratio of the Topcolor and QCD coupling
             constants at $R^{-1}=$ 10 TeV is taken to
             $g^2(R^{-1})/g_3^2(R^{-1}) = 4.5$.
             \label{fig-hatg}}
  \end{center}
\end{figure}

Once we specify $N_X$ and the Topcolor coupling at $R^{-1}$,
the RG flow of $\hat g^2$ is completely determined.
(See also Eq.~(\ref{g3-g}). )
We show typical RG flows in Fig.~\ref{fig-hatg}.
We used the following values of $\alpha_i (\equiv g_i^2/(4\pi))$
at $\mu=M_Z(=91.1876 \;{\rm GeV})$
as inputs of RGEs:~\cite{PDG}
\begin{eqnarray}
  \alpha_3(M_Z) &=& 0.1172, \label{qcd-mz} \\
  \alpha_2(M_Z) &=& 0.033822, \label{su2-mz} \\
  \alpha_Y(M_Z) &=& 0.010167 . \label{u1-mz}
\end{eqnarray}
We also note the value of $\alpha_3$ at $R^{-1}=$ 10 TeV evolved by
the 1-loop RGE,
\begin{equation}
    \alpha_3(\mbox{10 TeV}) = 0.07264.
\end{equation}

The beta function (\ref{rge_ED2}) of the bulk gauge coupling $\hat g$ 
implies the existence of the UVFP $g_*$ for $SU(3)_1$,
\begin{equation}
  g_*^2 \NDA = \frac{1}{-b_{\rm tc}}, \quad 
  \mbox{for} \quad N_X \leq 5.
\end{equation}
(This is different from the formula for the orbifold 
compactification by the factor $1/(1+2/\delta)$~\cite{Hashimoto:2000uk}.)
For $SU(3)_2$ the UVFP $g'_*$ is given by
\begin{equation}
  g'_*{}^2 \NDA = \frac{1}{-b'} \simeq 0.21. \label{UVFP2}
\end{equation}
We note here that the $U(1)_Y$ gauge interaction has the Landau pole $\Ly$
at which the gauge coupling constant diverges. 
Since our model includes $U(1)_Y$ in the bulk, 
we need to introduce a cutoff $\Lambda$ smaller than
the Landau pole $\Ly$.
The bulk gauge coupling $\hat g_Y (\mu)$ rapidly grows
due to the power-like behaviour of the running.
As a result, the Landau pole $\Ly$ is not so far from the compactification
scale $R^{-1}$.
Although $\Ly R \sim {\cal O}(10)$, the values of $\hat g (\mu \sim \Ly)$ 
and $\hat g' (\mu \sim \Ly)$ can be approximated by the UVFP values, 
$g_*$ and $g'_*$, respectively.
(See typical RG flows in Fig.~\ref{fig-hatg}.)
In order to realize the situation that the top condensation is 
favoured rather than the up and charm,
we thus require $g_* > g'_*$ and thereby obtain
\begin{equation}
  N_X \geq 3 .
\end{equation}
Whether or not the up and charm condensations are really suppressed 
is essentially determined by the model parameter
which yields $g'_*{}^2 \NDA \simeq 0.21$ as in Eq.~(\ref{UVFP2}).
In the next section, we will confirm that the suppression actually occurs 
and show that the energy region where only the top condensation takes
place does exist.

\section{Analysis of the ladder SD equation}
\label{sec5}

We explore the energy region where
only the top quark condenses while others do not (tMAC region).
Since our model explicitly breaks the six dimensional Lorentz symmetry,
it is not obvious whether or not the approach of the ladder SD equation
for the bulk fermion is appropriate.
Nevertheless we may adopt the ladder SD equation in six dimensions,
supposing the cutoff $\Lambda R \sim {\cal O}(10)$ is large enough.

The power-like running of the gauge couplings is crucial for
the analysis of the tMAC region.
Thus we should incorporate the running effects in the ladder SD equation.
Several methods have been applied to the phenomenology of the low-energy
QCD in four dimensions.
Simplest one is the Higashijima-Miransky approximation
in which the gauge coupling is replaced by~\cite{imp-simplest}
\begin{equation}
  g_3^2 \to g_3^2 (\max(-p^2,-q^2)),
\end{equation}
where $p$ and $q$ are external and loop momenta of the fermion, respectively.
However the Higashijima-Miransky approximation is inconsistent with
the axial Ward-Takahashi (WT) identity.
A natural choice is to take the argument of $g_3$ to
the gluon loop momentum $(p-q)$,
\begin{equation}
  g_3^2 \to g_3^2 (-(p-q)^2) .
\end{equation}
In this case, the ladder approximation can be consistent with
both of vector and axial WT identities~\cite{Kugo:1992pr}.
A demerit of the method is that the angular integration cannot be performed
analytically, i.e., the numerical calculation becomes complicated.
In Ref.~\cite{Aoki:1990aq}, it is shown that the approximation
\begin{equation}
  g_3^2 \to g_3^2 (-(p^2+q^2)) \label{g-ave}
\end{equation}
works well in four dimensions.
We may adopt Eq.~(\ref{g-ave}) even in extra dimensions.

Let us solve the ladder SD equation incorporating the running effects
by using the description,
\begin{equation}
  g_{6D}^2 \to \frac{\hat g^2 (-(p^2+q^2))}{-(p^2+q^2)}.
\end{equation}
For consistency with the vector Ward-Takahashi identity,
we choose the Landau gauge and then obtain the ladder
SD equation for the fermion mass function $B_f$ as follows:
\begin{eqnarray}
\lefteqn{ B_f (x) =} \nonumber \\
&& (D-1)\int_{R^{-2}}^{\Lambda^2} \!\!dy\,
 y^{D/2-1}\frac{B_f(y)}{y+B_f^2(y)}\frac{\kappa_f (x+y)}{x+y}
 K_B (x,y) , \nonumber \\
 \label{sd_b_imp}
\end{eqnarray}
with $f=t,b,u,c,\ell$, and $x \equiv -p^2$, and $y \equiv -q^2$, where
the kernel $K_B$ is given by~\cite{Hashimoto:2000uk}
\begin{equation}
  K_B (x,y) = \frac{1}{x}\left(1-\frac{y}{3x}\right) \theta (x-y) 
  + (x \leftrightarrow y) \quad  \mbox{for } D=6.
\end{equation}
We identified the infrared (IR) cutoff of the SD equation to
the compactification scale $R^{-1}$.
The binding strengths $\kappa_f$'s are
\begin{align}
  \kappa_t (\mu^2)&= C_F \hat g^2 (\mu) \NDA
                 \,+\, \frac{1}{9}\hat g_Y^2 (\mu) \NDA \label{k_t}, \\
  \kappa_b (\mu^2)&= C_F \hat g^2 (\mu) \NDA
                 \,-\, \frac{1}{18}\hat g_Y^2 (\mu) \NDA \label{k_b}, \\
  \kappa_{u,c} (\mu^2)&= C_F \hat g'{}^2 (\mu) \NDA
                 \,+\, \frac{1}{9}\hat g_Y^2 (\mu) \NDA \label{k_c}, \\
  \kappa_\ell (\mu^2)&= \phantom{C_F \hat g^2 (\mu) \NDA + \;\;\;\;}
                   \frac{1}{2} \hat g_Y^2 (\mu) \NDA \label{k_l},
\end{align}
for the top, bottom, up, charm, and lepton condensates, respectively.
The constant $C_F (= 4/3)$ is the quadratic Casimir of
the fundamental representation of $SU(3)$.
In the following analysis, we study these four channels.
The argument of $\kappa_f$ should be smaller than the Landau pole of
$U(1)_Y$, i.e.,
\begin{equation}
  \max(x+y) = 2\Lambda^2 < \Ly^2 .
\end{equation}

\begin{figure}[tbp]
  \begin{center}
    \resizebox{0.45\textwidth}{!}{
     \includegraphics{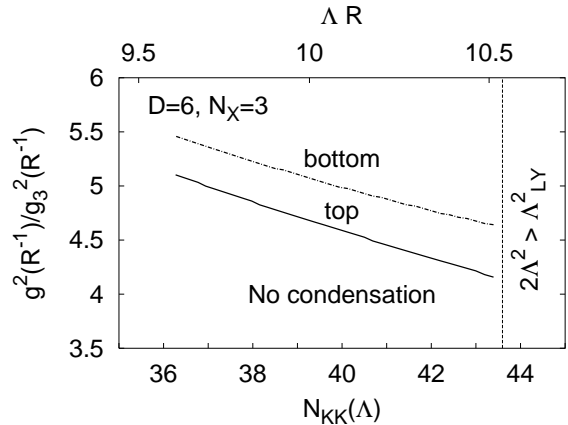}}
    \caption{The tMAC region for the model with $N_X=3$.
             The coupling constants of the Topcolor and QCD
             at the compactification scale $R^{-1}$ are represented as
             $g^2(R^{-1})$ and $g_3^2(R^{-1})$, respectively.
             In the ``top'' region only the top condensation occurs
             (tMAC region). In the ``bottom'' region
             the bottom quark condenses as well.
             No condensation takes place in the region of ``No condensation''.
             For $2\Lambda^2 > \Ly^2$ the argument of the gauge coupling
             in the kernel of the ladder SD equation exceeds the
             Landau pole of $U(1)_Y$.
             \label{fig1}}
  \end{center}
\end{figure}

We numerically solve the SD equation by using the iteration method, whose
details are described in Ref.~\cite{Hashimoto:2000uk}.
In the analysis, we fix the compactification scale $R^{-1}$ to 10 TeV.
For other values, the results are essentially unchanged.
We depict the result for the models with $N_X=3$ in Fig.~\ref{fig1}.
The ``top'' region in Fig.~\ref{fig1} corresponds to the tMAC.
In the ``bottom'' region, both of the top and bottom condensations
take place.
If we choose the ratio of the values of the Topcolor and QCD couplings
at $R^{-1}=$ 10 TeV to
$g^2(R^{-1})/g_3^2(R^{-1}) \sim 4.2$--$4.6$,
the tMAC region is $\Lambda R \sim 10$--$10.5$.
In the region it turns out that the up- and
charm-condensations do not occur.
For $\Lambda R > 10.5$ the lepton condensation is favoured.

Similarly, the tMAC regions are also found for models with $N_X=4,5$.
However the regions become narrower:
for $g^2(R^{-1})/g_3^2(R^{-1}) \sim 2.1$--$2.3$, $\Lambda R \sim 10.2$--$10.5$,
$(N_X=4)$;
for $g^2(R^{-1})/g_3^2(R^{-1}) \sim 1.3$--$1.4$, $\Lambda R \sim 10.3$--$10.5$,
$(N_X=5)$.

\section{Summary and discussions}
\label{summary}

We studied the Topcolor model in the six dimensional bulk.
We assigned the nontrivial BC's to the Topcolor gauge fields
so that the Topcolor is broken down on the boundaries.
As a three generation model
we considered the model whose charge assignments are shown in
Table~\ref{tab1}.
Since the top and bottom quarks have the Topcolor charge while
the other quarks do not in the model, the up and charm condensations
are unlikely to occur.
When the bulk $U(1)_Y$ interaction is sufficiently strong,
the bottom condensation is also suppressed.
In this way, we can expect that only the top quark condenses,
which is required for a viable model.
In order to demonstrate the existence of such a situation,
we analyzed the ladder SD equation including the RGE effects of
the bulk gauge couplings.
We then found that the situation can be realized in the ``top'' region
shown in Fig.~\ref{fig1}, which is the result for the model
with three extra (heavy) vector-like fermions having the Topcolor charge,
i.e., $N_X=3$.
For example, when the ratio of the couplings of Topcolor and QCD is
taken to $g^2(R^{-1})/g_3^2(R^{-1}) \sim 4.2$--$4.6$ with
$R^{-1}(\approx \mbox{10 TeV})$ being the compactification scale,
the cutoff $\Lambda$ should be $\Lambda R \sim 10$--$10.5$.
The models with $N_X=4,5$ may be possible as well.

The electroweak gauge sector of the model is the same as the universal
extra dimension model~\cite{Appelquist:2000nn}.
The compactification scale $R^{-1}$ is severely constrained by
the LEP precision data~\cite{PDG}.
Since the KK modes of bulk fermions are vector-like, the constraint from
the $S$ parameter is suppressed,
$S \approx 10^{-2} \sum_{j,k}\frac{m_t^2}{M_{j,k}^2}$.
Hence the $T$-parameter constraint is essential.
We may estimate the $T$-parameter as in Ref.~\cite{Appelquist:2000nn},
\begin{equation}
  T \approx 0.76 \sum_{j,k}\frac{m_t^2}{M_{j,k}^2} ,
\end{equation}
where we neglected ${\cal O}(m_t^4/M_{j,k}^4)$ contributions.
When we take $\max(M_{j,k})=\Lambda$ or $\sqrt{2}\Lambda$ with
$\Lambda \sim (10\mbox{--}10.5)R^{-1}$, the estimate of the $T$-parameter is
\begin{equation}
T \approx (4\mbox{--}5)\times
10^{-2}\frac{(1\;\mbox{TeV}^2)}{R^{-2}(\mbox{TeV}^2)}.
\end{equation}
The current constraint $T < 0.02$ at $95\%$ C.L. with the Higgs boson mass
$m_H=117$ GeV~\cite{PDG} yields $R^{-1} > 1.4$--$1.6$~TeV.
For larger $m_H$ the lower bound of $R^{-1}$ gets smaller.
For the reference value $R^{-1}=10$ TeV,
we can expect that the contributions of KK modes to the $T$-parameter is
negligibly small,
even if we take into account errors arising from nonperturbative effects.

If the compactification scale $R^{-1}$ is larger than a several TeV, 
the KK modes will not be observed directly in LHC.
In such a case, the Higgs production via the gluon fusion process 
may become important in order to discriminate the present model from
the SM.~\cite{Hashimoto:2002cy}

The mass of the top quark may be predicted larger than 
the experimental value, because a stronger gauge coupling than QCD
was used in the model and the top-Yukawa couplings is determined
through the infrared value of the gauge coupling from the viewpoint of 
RGEs~\cite{Bardeen:1989ds}.
If so, the top-seesaw mechanism~\cite{Dobrescu:1997nm} is helpful. 
When we apply the top-seesaw scheme, we need to introduce a heavy fermion
with the same charge assignment as $t_R$.

Our approach is very sensitive to the cutoff, so that
the UV completion by theory space~\cite{Arkani-Hamed:2001ca,Hill:2000mu}
may be required.

\section*{Acknowledgments}

The author thanks M. Tanabashi for useful comments concerning the BC's 
for gauge scalars.

\end{document}